\documentclass[%
twocolumn,
 amsmath,amssymb,
aps,
pra,
]{revtex4-1}
\usepackage{listings}
\usepackage{graphicx}
\usepackage{dcolumn}
\usepackage{bm}
\usepackage{natbib}
\usepackage[colorlinks,citecolor=red,linkcolor=blue,urlcolor=blue]{hyperref}
\usepackage{capt-of}
\usepackage{hhline}
\usepackage{float}
\usepackage{mhchem}
\usepackage[nodisplayskipstretch]{setspace}
\usepackage{calrsfs}
	\DeclareMathAlphabet{\pazocal}{OMS}{zplm}{m}{n}
\usepackage{physics}
\usepackage{appendix}
\usepackage[export]{adjustbox}
\usepackage[normalem]{ulem}  

\newcommand{\bk}{\bold{k}}

\newcommand{\squeezeD}[3]{\left\langle #1\middle| #2\middle| #3\right\rangle}

\usepackage[dvipsnames]{xcolor}
\definecolor{pink}{rgb}{0.858, 0.188, 0.478}

\usepackage{lineno,blindtext}

\begin{document}


\title{Weyl triplons in \ce{SrCu2(BO3)2}}

\author{Dhiman Bhowmick\href{https://orcid.org/0000-0001-7057-1608}{\includegraphics[scale=0.12]{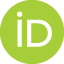}}}
\author{Pinaki Sengupta\href{https://orcid.org/0000-0003-3312-2760}{\includegraphics[scale=0.12]{orcid.png}}}%
\affiliation{
 School of Physical and Mathematical Sciences, Nanyang Technological University, 21 Nanyang Link, Singapore 637371, Singapore \\
}

\date{\today}

\begin{abstract}

We propose that Weyl triplons are expected to appear in the low 
energy magnetic excitations in the canonical Shastry-Sutherland compound, 
\ce{SrCu2(BO3)2}, a quasi-2D quantum magnet. Our results show that when
a minimal, realistic inter-layer coupling is added to the well-established microscopic
model describing the excitation spectrum of the individual layers,
the Dirac points that appear in the zero-field triplon spectrum of the 2D model
split into two pairs of Weyl points along the $k_z$ direction. Varying
the strength of the inter-layer DM-interaction and applying a
small longitudinal magnetic field results in a range of band-topological
transitions accompanied by changing numbers of Weyl points. We propose inelastic neutron scattering along with thermal Hall effect as the experimental techniques to detect the presence of Weyl node in the triplon spectrum of this material.
 We show that the logarithmic divergence in the second derivative in thermal Hall conductance near phase transition from regime Weyl points to a regime with topologically gapped bands as well as a finite slope in the thermal Hall conductance as a function of magnetic field at zero magnetic field are promising evidence for the presence of Weyl triplons.


\end{abstract}

\pacs{Valid PACS appear here}
\maketitle


\section{\label{sec1}Introduction}

The successful detection of Weyl fermions in \ce{TaAs}\,\cite{TaAs}, following 
theoretical prediction of the same\,\cite{TaAs_Theory1,TaAs_Theory2}, marks one of the latest 
milestones in the study of topological phases of matter, currently
the most active frontier in Condensed Matter Physics\,\cite{WSM1,WSM2,WSM3,WSM4,WSM5,WSM6,WSM7,WSM10,WSM11}. Weyl fermions
are massless, linearly dispersing quasiparticles with finite chirality,
first proposed as solutions to massless Dirac equation in relativistic
particle physics\,\cite{Weyl}.  
In condensed 
matter systems, non-relativistic analog of Weyl quasiparticles emerge at 
linear crossing of non-degenerate, topologically protected bands 
in three dimensional reciprocal space. Interest in these special band
crossings have increased since they act as sources of Berry flux and
impart topological character to the associated energy bands. 
Weyl nodes appear in pairs with opposite chirality and can be separated in
momentum space in systems with broken time reversal\,\cite{WSM4,WSM5,WSM6,WSM7} or inversion symmetry\,\cite{WSM10,WSM11} or both\,\cite{WSM12}.

Like many other topological features, the appearance of Weyl points is governed by the geometry of the band structure
and symmetries of the Hamiltonian and lattice, and independent of the quantum statistics. As such, it is possible to
observe bosonic analogs of Weyl points. This has already been achieved in
artificially designed photonic\,\cite{photonics1,photonics2,photonics3,photonics4} and phononic crystals\,\cite{phonon1,phonon2}, and proposed for
magnons\,\cite{WM1,WM2,WM3,WM4,WM5,WM6,KagomeAntiFerromagnet_WeylMagnon}. 
 Weyl points with toplogical charges $\pm 2$ are found in the phonon spectra\,\cite{DoublePhonon1,DoublePhonon2} and excitation spectra in phononic\,\cite{DoublePhononic1,DoublePhononic2} and photonic crystals\,\cite{DoublePhotonic1,DoublePhotonic2} which have no counterpart in high-energy physics. 
But no such unconventional Weyl points have been reported in electronic or magnonic systems.
Our results reveal that magnetic excitations in the real quantum magnet \ce{SrCu2(BO3)2} is a promising platform to realise this unique doubly charged Weyl points in magnetic excitations for the first time.  
The ground state magnetic-properties of the quasi-two dimensional quantum magnet \ce{SrCu2(BO3)2} is well described by the canonical Shastry-Sutherland model\,\cite{Shastry1981,Kageyama1999,Kodama2002,Gaulin2004,Sebastian2008,Joliquer2001}. The material also features  additional symmetry allowed Dzyaloshinskii–Moriya(DM)-interactions\,\cite{Nojiri1999,Experiment1}, although it
is sufficiently weak to have pronounced effects on the ground state magentization. However, the DM-interaction has dramatic effects on magnetic excitations.
Romhányi et al. studied low-lying magnetic excitations in the extended model and established that the triplon excitations acquire topological characteristics in the presence of DM-interactions\,\cite{triplon0,triplon}.
Malki et al. further extended the study showing several band-topological phase transitions in presence of a more general uniform magnetic field\,\cite{triplon2}.
In contrast to past theoretical studies, we study the topological properties of the magnetic excitations of the three-dimensional Shastry-Sutherland model with additional inter-layer symmetry-allowed spin-interactions.

Quantum magnets are particularly promising since they have for long
been a versatile platform to realise
bosonic analogs of novel fermionic phases. The wide range of available
quantum magnets with different lattice geometries and the ability to tune
their properties readily by external magnetic field make them ideal testbed
for realising bosonic topological phases\,\cite{TopologicalMagnon,TopologicalMagnon2,TopologicalMagnon3,TopologicalMagnon4,TopologicalMagnon5,TopologicalMagnon6,TopologicalMagnon7,new1,new2,new3,new4,new5, MagnonPolaron1,MagnonPolaron2,MagnonPolaron3,MagnonPolaron4,WM1,WM2,WM3,WM4,WM5,WM6}. However,
despite theoretical predictions, experimental observation of Weyl points in magnetic excitations
have remained elusive. In this work,
we present evidence for the existence of Weyl triplons in the geometrically
frustrated Shastry Sutherland compound, \ce{SrCu2(BO3)2}. In contrast to
previous studies that considered idealized Hamiltonians for multiple families
of quantum magnets\,\cite{WM1,WM2,WM3,WM5}, we focus on a realistic microscopic Hamiltonian of the
extensively studied geometrically frustrated quantum magnet, \ce{SrCu2(BO3)2}\,\cite{triplon,triplon2,triplon3}.

Previous theoretical studies have shown that stacking two-dimensional quantum magnets with topological  magnon bands\,\cite{TopologicalMagnon,KagomeFerromagnet_TopologicalMagnon} may give rise to topological Weyl magnons\,\cite{KagomeFerromagnet_WeylMagnon,WM6}.
In this study we show that geometrically frustrated quasi-two-dimensional Shastry-Sutherland material \ce{SrCu2(BO3)2}, consisting of weakly coupled Cu-O planes, is a promising candidate for observing Weyl triplons within realistic parameter ranges. We use a microscopic model with experimentally determined Hamiltonian parameters\,\cite{ESR} that has been demonstrated to reproduce faithfully 
the observed behaviour of the material\,\cite{triplon3}.

Our paper is structured as follows. In Sec.\,\ref{sec2}, we introduce the microscopic model and the derivation of the triplon Hamiltonian.
Afterwards, we present our results in Sec.\,\ref{sec3} which is further subdivided into three subsections. 
In Sec.\,\ref{sec3a} we show the presence of topological Weyl triplons and band-topological phase diagrams. 
In Sec.\ref{sec3b}, the presence of surface arcs and surface states are shown as a signature of topological nature of Weyl triplon.
Finally, in Sec.\,\ref{sec3c}, we calculate the associated triplon thermal Hall conductance and provided a detailed analysis of the nature of thermal Hall conductance and its' derivatives as a function of magnetic field to experimentally detect Weyl triplons.
The principal findings are summarized and comparative discussion with respect to previous neutron scattering experiment is provided in section Sec.\,\ref{sec4}.

\section{\label{sec2} Model}
\subsection{Microscopic spin model.}


\begin{figure}[tb]
\includegraphics[width=0.47\textwidth]{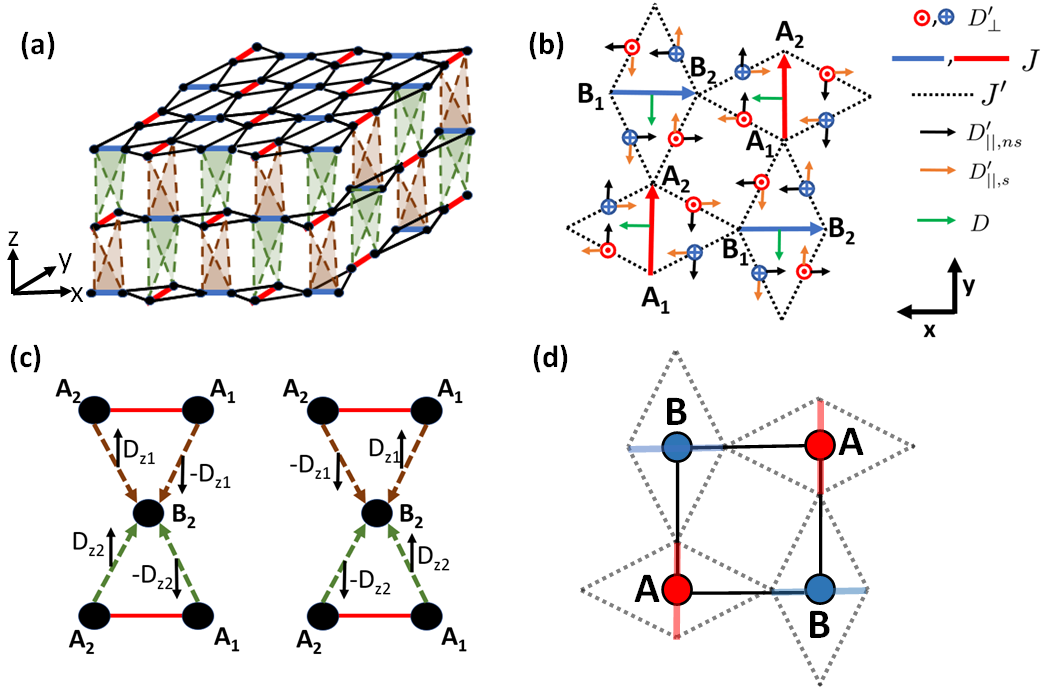} 
\caption{(a) The 3D-schematic lattice structure of compound \ce{SrCu2(BO3)2}. The red-bonds are dimer-A and blue-bonds are dimer-B. The black bonds are axial bonds. In real material the dimer-A and dimer-B bonds are out of plane of 2D-layer in a way that some interlayer bonds are shorter than the other. As a consequence, two different kinds of inter-layer bonds originate which is depicted via green or brown dotted lines. (b) The intralayer Heisenberg and DM-interactions. (c) The interlayer DM-interactions. (d) The effective square lattice after bond-operator transformation on the two-dimensional Shastry-Sutherland (SS) lattice.}
\label{lattice}
\end{figure}

Figure\,\ref{lattice}(a) illustrates the three dimensional arrangements of \ce{Cu}$^{2+}$-ions of \ce{SrCu2(BO3)2} as stacked planes of Shastry-Sutherland (SS) lattice. The Hamiltonian of this spin-half system is given by,
\begin{align}
\pazocal{H}=&J\sum_{\left\langle i,j \right\rangle,l} \bold{S}_{i,l}\cdot\bold{S}_{j,l} + J' \sum_{\left\langle\left\langle i,j \right\rangle\right\rangle,l} \bold{S}_{i,l}\cdot\bold{S}_{j,l} \nonumber\\
  + \bold{D}&\cdot\sum_{\left\langle i,j \right\rangle,l} \left(\bold{S}_{i,l}\times\bold{S}_{j,l}\right)+ \bold{D'}\cdot\sum_{\left\langle\left\langle i,j \right\rangle\right\rangle,l} \left(\bold{S}_{i,l}\times\bold{S}_{j,l}\right)
  \nonumber\\
  +J_z^b &\sum_{\substack{\left\langle i,j\right\rangle_b,\\ \left\langle l,l'\right\rangle}} \bold{S}_{i,l} \cdot\bold{S}_{j,l'}+\bold{D}_z^b\cdot\sum_{\substack{\left\langle i,j\right\rangle_b ,\\ \left\langle l,l'\right\rangle}} \left(\bold{S}_{i,l}\times\bold{S}_{j,l'}\right)
  \nonumber \\
  +J_z^g &\sum_{\substack{\left\langle i,j\right\rangle_g,\\ \left\langle l,l'\right\rangle}} \bold{S}_{i,l} \cdot\bold{S}_{j,l'}+\bold{D}_z^g\cdot\sum_{\substack{\left\langle i,j\right\rangle_g,\\ \left\langle l,l'\right\rangle}} \left(\bold{S}_{i,l}\times\bold{S}_{j,l'}\right)
  \nonumber\\
  -g_z h_z & \sum_{i,l} S_{i,l}^z,
  \label{eq1}
\end{align}
where, 
$\left\langle i,j \right\rangle$ and $\left\langle\left\langle i,j\right\rangle\right\rangle$ denote the summation over the sites belonging to intra-dimer (dimer-A or red-bond and dimer-B or blue-bond in Fig.\,\ref{lattice}(a) ) and inter-dimer bonds (or axial-bonds or black-bonds in Fig.\,\ref{lattice}(a) ) respectively within a SS-layer and
$\left\langle l,l'\right\rangle$ denotes pairs of adjacent SS-layers. $\left\langle i,j\right\rangle_b$ and $\left\langle i,j\right\rangle_g$ denote the nearest neighbour inter-layer blue and green bonds
respectively as in figure Fig.\,\ref{lattice}(a) and Fig.\,\ref{lattice}(c). 
The first four terms describe the intra-layer coupling terms and are depicted in 
Fig.\,\ref{lattice}(b), where $J$ and $J'$ are the intra-dimer and inter-dimer Heisenberg terms respectively. We include DM-interactions that are symmetry allowed for \ce{SrCu2(BO3)2} at 
temperatures below $395$K\,\cite{structure1,structure2,triplon} in its low-symmetry
phase.
 $\bold{D}$ and  $\bold{D}'$ denote the intra-dimer and 
inter-dimer Dzyaloshinskii–Moriya(DM)-interactions respectively. 
 $D_\perp^\prime$, $D^\prime_{||,s}$ (staggered) and $D^\prime_{||,ns}$ (non-staggered) are the components of DMI $\bold{D}'$ as shown in figure Fig.\,\ref{lattice}(b). 
$J_z^b$ and $D_z^b$ ($J_z^g$ and $D_z^g$) are the inter-layer Heisenberg terms and DM-interactions on the brown (green) dotted bonds in Fig.\,\ref{lattice}(a) and Fig.\,\ref{lattice}(c) respectively. 
The last term is a Zeeman coupling of the spins 
with a magnetic field where $h_z$ is the magnetic field perpendicular to the SS-layer and $g_z$ is the g-factor.


The 2D Hamiltonian describing the magnetic properties of each layer have been
extensively studied in the past and the nature
of triplon excitations above dimerized ground state and their topological
characters delineated using the bond operator formalism\,\cite{triplon,triplon3} and higher-order series expansion\,\cite{triplon2}.
We study the system with additional physically realistic interlayer Heisenberg and DM-interactions. The interlayer DM-interaction shown in Fig.\,\ref{lattice}(c), is taken in the z-direction. Although allowed by symmetry, the transverse components of the inter-layer DM-interaction are neglected, since their contribution to the low energy physics of the magnetic system is found to be negligible compared to that from $D^\prime_{||,ns}$. For simplicity, we assume $D_{z2}\approx D_{z1}=D_z$. The presence of interlayer DM-interaction drives a variety of topological phases in the system.

\subsection{\label{Triplon Picture} Triplon Hamiltonian}
If $J'\lesssim 0.7J$ the ground state of the canonical Shastry-Sutherland model (first two terms in Eq.(\,\ref{eq1})) is a dimer product state\,\cite{PhaseTransition1,PhaseTransition2}, whereas for the material \ce{SrCu2(BO3)2},experimentally  measured $J'$ lies in between $0.6J$ to $0.68J$ \,\cite{Experiment1,Experiment2,Experiment3}, indicating the material possesses a low temperature dimer ground state.
The inter-layer Heisenberg exchange interaction is sufficiently small so as to preserve the dimer ground state of the material\,\cite{InterLayerCoupling,InterLayerCoupling2}.

 The dimers consist of singlets,  $\ket{s}=\left(\ket{\uparrow\downarrow}-\ket{\downarrow\uparrow}\right)/\sqrt{2}$,  on each diagonal bonds (red and blue bonds in Fig.\,\ref{lattice}(a),(b) ) of a Shastry-Sutherland lattice. The lowest magnetic excitations above the ground state consist of triplets: $\ket{t^x}=i\left(\ket{\uparrow\uparrow}-\ket{\downarrow\downarrow}\right)/\sqrt{2}$, $\ket{t^y}=\left(\ket{\uparrow\uparrow}+\ket{\downarrow\downarrow}\right)/\sqrt{2}$, $\ket{t^z}=-i\left(\ket{\uparrow\downarrow}+\ket{\downarrow\uparrow}\right)/\sqrt{2}$. The widely used bond operator formalism is ideally suited to investigate the properties of low temperature excitation above the ground state in dimerized systems\,\cite{triplon0, triplon}. In this formalism, the product state of singlets form the vacuum and the local quasi-particle triplon excitations are described by a bosonic Hamiltonian. The effective lattice for the triplons is a square lattice consisting of two different sub-lattices as shown in Fig.\,\ref{lattice}(d). The following transformations relate the spin operators to the singlet and triplon creation/annihilation operators\,\cite{BondOperator1,BondOperator2}: \begin{eqnarray}
    S^\mu_{j,1}&=&
    \frac{i}{2}\left(t_{j}^{\mu\dagger} s_j
    -s_j^\dagger t_{j}^\mu\right)
    -\frac{i}{2}\epsilon_{\mu\nu\eta} t_{j}^{\nu\dagger} t_{j}^\eta,\nonumber\\
    S^\mu_{j,2}&=&
    -\frac{i}{2}\left(t_{j}^{\mu\dagger} s_j
    -s_j^\dagger t_{j}^\mu\right)
    -\frac{i}{2}\epsilon_{\mu\nu\eta} t_{j}^{\nu\dagger} t_{j}^\eta,
\end{eqnarray}
where $\mu = x,y,z$, $(j,1)$ and $(j,2)$ are the two spins connected by the diagonal bond $j$ (red and blue bonds in Fig.\,\ref{lattice}(a),(b) ) and $\epsilon_{\mu\nu\eta}$ is the Levi-Civita symbol for cyclic permutation of $\{x,y,z\}$. Using the above transformation, one can derive the effective triplon Hamiltonian from Eq.(\,\ref{eq1}) by assuming $s_j=s_j^\dagger=\langle s_j\rangle \approx 1$, that is, the ground state is a condensation of the singlet states. In the simplest mean field approximation that we shall use here, one retains only terms up to bilinear in the triplon operators, yielding a tight binding model of triplons on an effective square lattice where each lattice site corresponds to a single dimer and the bonds represent the coordination between the neighbouring dimers (see Fig.\,\ref{lattice}(d)).  

In the presence of the small on-dimer DM-interaction $\bold{D}$, the  ground state retains its dimer-product character to the lowest order in perturbation, but the states on the two diagonal bonds in the unit cell (fig.\,\ref{lattice}(b)) are rendered inequivalent. The local Hilbert space on each diagonal consists of superpositions of singlets and triplets of the constituent spins that can be represented as (the subscripts $A$ and $B$ denote the dimer bonds $A$ and $B$ respectively):  
\begin{equation}
\begin{pmatrix}
\ket{\tilde{s}_A} \\
\ket{\tilde{t}^x_A} \\
\ket{\tilde{t}^y_A} \\
\ket{\tilde{t}^z_A} \\
\end{pmatrix}=
\begin{pmatrix}
1 & -\alpha & 0 & 0\\
\alpha & 1 & 0 & 0\\
0 & 0 & 1 & 0\\
0 & 0 & 0 & 1
\end{pmatrix}
\begin{pmatrix}
\ket{s_A} \\
\ket{t^x_A} \\
\ket{t^y_A} \\
\ket{t^z_A} \\
\end{pmatrix},
\end{equation}

\begin{equation}
\begin{pmatrix}
\ket{\tilde{s}_B} \\
\ket{\tilde{t}^x_B} \\
\ket{\tilde{t}^y_B} \\
\ket{\tilde{t}^z_B} \\
\end{pmatrix}=
\begin{pmatrix}
1 & 0 & \alpha & 0\\
0 & 1 & 0 & 0\\
-\alpha & 0 & 1 & 0\\
0 & 0 & 0 & 1
\end{pmatrix}
\begin{pmatrix}
\ket{s_B} \\
\ket{t^x_B} \\
\ket{t^y_B} \\
\ket{t^z_B} \\
\end{pmatrix},
\end{equation}
where, $\alpha\approx\frac{|D|}{2J}+\pazocal{O}(\frac{|D|^3}{J^3})$ and $\ket{\tilde{s}_A}$ and $\ket{\tilde{s}_B}$ are the new lowest energy state on diagonal $A$ and $B$ respectively. We use the bond operator formalism decribed above to derive the effective triplon Hamiltonian, retaining terms up to liner order in $\alpha$.
 The low energy triplon Hamiltonian is further transformed using unitary transformation, such that the two sub-lattices in the effective square lattice in figure Fig.\,\ref{lattice}(d) become equivalent\,\cite{triplon}. The triplon-Hamiltonian after neglecting the terms of order of $\alpha^2$ (since $\alpha\ll 1$) is given by,


\begin{eqnarray}
\pazocal{H}&=&J\sum_{j}\sum_{\mu=x,y} \tilde{t}^{\mu\dagger}_j\tilde{t}^\mu_j+i g_z h_z\sum_j\left[\tilde{t}^{x\dagger}_j\tilde{t}^y_j-\tilde{t}^{y\dagger}_j\tilde{t}^x_j\right]\nonumber \\
&-&\frac{iD'_\perp}{2} \sum_j\sum_{\delta=\hat{x},\hat{y}} \left[\tilde{t}^{y\dagger}_j \tilde{t}^x_{j+\delta} +\tilde{t}^{y\dagger}_j\tilde{t}^x_{j+\delta} - \text{h.c.}\right]\nonumber\\
&+&\frac{i\tilde{D}'_{||,s}}{2}\sum_j\left[\tilde{t}^{z\dagger}_{j+\hat{x}}\tilde{t}^y_j+\tilde{t}^{y\dagger}_{j+\hat{x}}\tilde{t}^z_j-\text{h.c.}\right.\nonumber\\
& &\left. \quad\quad\quad\quad -\tilde{t}^{z\dagger}_{j+\hat{y}}\tilde{t}^x_j-\tilde{t}^{x\dagger}_{j+\hat{y}}\tilde{t}^z_j - \text{h.c.}\right]\nonumber\\
&-&iD_z\sum_j \left[\tilde{t}^{y\dagger}_{j+\hat{z}}\tilde{t}^x_j+\tilde{t}^{y\dagger}_{j}\tilde{t}^x_{j+\hat{z}}-\text{h.c.}\right].
\label{eq:triplon_H}
\end{eqnarray}
Here the label $j$ denotes the positions of the diagonal bonds and $\{\hat{x},\hat{y},\hat{z}\}$ denote the nearest neighbor along
the three principal axes of the effective square lattice of dimers (see Fig.\,\ref{lattice}(d)). The terms with $J_z$, $D'_{||,ns}$ do not contribute to the energy to $\pazocal{O}(\alpha)$ due frustrated orthogonal dimer arrangement. The renormalized DM-interaction $\tilde{D}'_{||,s}$ is defined as $\tilde{D}'_{||,s}=D'_{||,s}-\frac{|D|J'}{2J}$.
The interplay between the DM-interactions $D_z$ and $D'_\perp$ generates different kinds of Weyl triplons in the system. In the following sub-section, we discuss the triplon energy spectrum and the emergence of toplogical Weyl triplons in the system.

The mean field Hamiltonian Eq.(\,\ref{eq:triplon_H}) possesses translational symmetry and the crystal-momentum is a conserved quantum number for the system. Hence one can derive the triplon band structure from solving the momentum space Hamiltonian, obtained from the Fourier transformation of the triplon operators,
\begin{equation}
\pazocal{H}=\sum_\bk\sum_{\mu,\nu=x,y,z} \hat{\tilde{t}}^\dagger_{\mu,\bk} M_{\mu\nu}(\bk) \hat{\tilde{t}}_{\nu,\bk},
\end{equation}
where the matrix $M(\bk)$ is given by,
\begin{align}
M(\bk)&=\begin{pmatrix}
J & i\tilde{h}_z & \tilde{D}'_{||}\gamma_2 \\
-i\tilde{h}_z & J & -\tilde{D}'_{||}\gamma_1 \\
\tilde{D}'_{||}\gamma_2 & -\tilde{D}'_{||}\gamma_1 & J
\end{pmatrix}\nonumber\\
&=J\mathbf{I}-\tilde{h}_z\lambda_2+\tilde{D}'_{||}\gamma_2 \lambda_4-\tilde{D}'_{||}\gamma_1 \lambda_6
\label{Matrix}
\end{align}
where, $\gamma_1=\sin(k_x)$, $\gamma_2=\sin(k_y)$, $\gamma_3=\frac{1}{2}(\cos(k_x)+\cos(k_y))$, $\gamma_4=\cos(k_z)$ and $\tilde{h}_z=g_zh_z+2D'_\perp\gamma_3+2D_z\gamma_4$. Moreover $\lambda_2,\,\lambda_4$ and $\lambda_6$ are Gell-Mann matrices.

\section{\label{sec3} Results}
\subsection{\label{sec3a} Topological Weyl triplons.}
We study the model fixing the parameters $J=722$ GHz, $\left|\tilde{D}'_{||}\right|=20$ GHz, $D'_\perp=-21$ GHz and $g_z=2.28$\,\cite{triplon, ESR} and varying the parameters $h_z$ and $D_z$. At a fixed $\bk$-point in momentum space the matrix Eq.\,\ref{Matrix} has three eigenvalues, $J$, $J+\frac{|\bold{d}(\bk)|}{2}$ and $J-\frac{|\bold{d}(\bk)|}{2}$, where $\bold{d}(\bk)=[\tilde{D}'_{||}\gamma_1,\tilde{D}'_{||}\gamma_2,-\tilde{h}_z]$.
Thus at low energies, the system has three different triplon bands that can cross at the high-symmetry points on the $k_x$-$k_y$ plane: $(\pi,0)$, $(0,\pi)$, $(0,0)$, $(\pi,\pi)$. The Weyl points at the high symmetry points are triply degenerate, which has no equivalence in high energy physics, because the qusiparticle excitation triplons in this system do not follow the Poincar\'e symmetry\,\cite{BeyondWeyl,BeyondWeylMagnon}.

\begin{widetext}

\begin{figure}[htb]
\includegraphics[width=\textwidth]{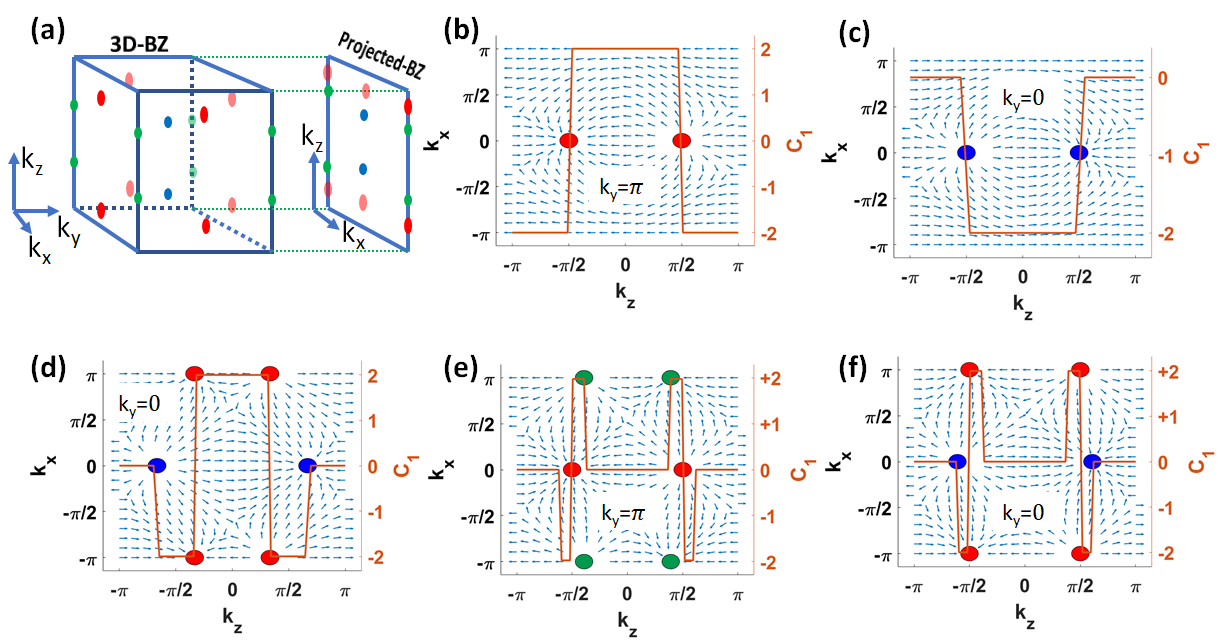} 
\caption{ (a) A schematic representation of all possible Weyl points in the first Brillouin zone. The color coding is described in the main text. (b)-(f) The direction of Berry curvature (blue-arrows) in $k_x-k_z$ plane for a fixed $k_y$ and Chern number of lowest triplon band (shown in red-line with right-vertical axis and calculated for bands in $k_x-k_y$-plane fixing $k_z$) plotted as function of $k_z$ for parameters (b) $D_z=D'_\perp/2$, $h_z=0$,  (c) $D_z=D'_\perp/2$, $h_z=h_c$, (d) $D_z=D'_\perp$, $h_z=h_c/2$, (e) $D_z=3D'_\perp$, $h_z=0$, (f) $D_z=3D'_\perp$, $h_z=0$. Where, $h_c=\frac{2\left|D'_\perp\right|}{g_z}=1.316$ T is band-topological phase transition point in absence of interlayer DMI $D_z$ and $D'_\perp=-21$ GHz.}
\label{Weyl}
\end{figure}

\end{widetext}

A schematic illustration of different types of Weyl points in the Brillouin zone(BZ) is shown in Fig.\,\ref{Weyl}(a). The red-dots illustrate the Weyl points at positions $\left(0,\pi,k_{z1}\right)$ and $\left(\pi,0,k_{z1}\right)$, where $k_{z1}=\cos^{-1}\left(-\frac{h_zg_z}{2D_z}\right)$. The blue dotes denote the Weyl points at position $(0,0,k_{z2})$, where $k_{z2}=\cos^{-1}\left(-\frac{h_zg_z+2D'_\perp}{2D_z}\right)$. Finally, the green-points are Weyl points at position $(\pi,\pi,k_{z3})$, where $k_{z3}=\cos^{-1}\left(\frac{2D'_\perp-h_zg_z}{2D_z}\right)$. 

 To verify the band crossings are topological Weyl points, we plot the direction of Berry curvature and change in Chern number within the first-BZ in Fig.\,\ref{Weyl}(b)-(f), for different parameter regions.
We note that the Chern number is defined strictly for a two-dimensional band; in this study, the Chern number is defined for the lower band in two-dimensional $k_x-k_y$ planes at a fixed $k_z$-value in the 3D Brillouin zone and it is defined for $n$-th band as,
\begin{equation}
    C_n(k_z)=\frac{1}{2\pi}\int^\pi_{-\pi}\int^\pi_{-\pi} dk_x dk_y \Omega_n^z(\bk),
\end{equation}
where, $\Omega^z_n(\bk)$ is $z$-component of Berry curvature of n-th band (n=1 denotes lowest band) at $\bk$-point in Brillouin zone which is given by,
\begin{widetext}
\begin{equation}
    \Omega^z_n(\bk) = i\sum_{m\neq n} 
    \frac{\squeezeD{m(\bk)}{\frac{\partial \pazocal{H}}{\partial k_x}}{n(\bk)} 
    \squeezeD{n(\bk)}{\frac{\partial \pazocal{H}}{\partial k_y}}{m(\bk)}-(k_x \leftrightarrow k_y)}{(E_n(\bk)-E_m(\bk))^2},
    \label{BerryCurvature}
\end{equation}
\end{widetext}
where $E_n(\bk)$ and $\ket{n(\bk)}$ denote the eigenvalue and eigenstate of $n$-th band at $\bk$-point in Brillouin zone respectively.
Three-band tight binding models have previously been studied for two-dimensional systems and found to have topologically gapped bands with Chern numbers of three bands $(+c,0,-c)$ or $(+c,-2c,+c)$ with $c\in \mathbb{Z}$\,\cite{MagnonPolaron3,triplon,Extra1,Extra2,Extra3}.
In this study, the calculated Chern numbers $C_n(k_z)$ of the three gapped bands at a fixed $k_z$-plane are found to be $(2c,0,-2c)$ with $c=\pm 1$ or $0$ which is similar to the two-dimensional counterpart of the model studied in the reference Ref.\,\cite{triplon}.
Weyl points are band-topological transition points in three dimensional Brillouin zone resulting in change in Chern numbers $C_n(k_z)$.
It is found that the Chern number changes by $\pm 2$ for the Weyl points present at $(0,0,\pm k_{z2})$ and $(\pi,\pi,\pm k_{z3})$ which indicates that the monopole charge associated with these Weyl points are $\pm 2$. At momenta $(0,\pi,\pm k_{z1})$ and $(\pi,0,\pm k_{z1})$, the Chern number changes by $\pm 4$  due to the joint contributions from the Weyl points , each of which carries a monopole charge of $\pm 2$.
Thus all the Weyl points are doubly charged Weyl points in the system.

\begin{widetext}

\begin{figure}[htb]
\includegraphics[width=\textwidth]{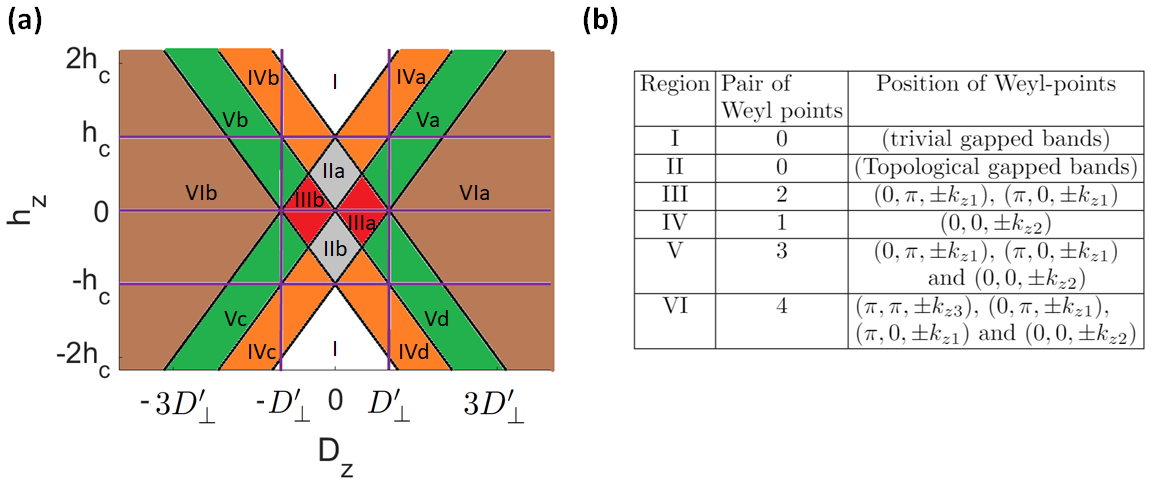} 
\caption{ (a) Different regimes of topological triplon bands are defined based on the number of Weyl points and their positions in the BZ. The subdivision of the regions a, b, c, d denote the changes in the charge of Weyl points. (b) The number and positions of Weyl points at different regions of parameter space are provided in the table. Here, $h_c=\frac{2\left|D'_\perp\right|}{g_z}=1.316$ T is band-topological phase transition point in absence of interlayer DMI $D_z$ and $D'_\perp=-21$ GHz.}
\label{Phase}
\end{figure}

\end{widetext}

Based on the number of Weyl points and their positions in the $k_x$-$k_y$ plane, we divide the $h_z$-$D_z$ parameter space into several regions as shown in Fig.\,\ref{Phase}(a). Regions I and II feature no Weyl points.  The triplon bands in region-II are topological in nature; Chern number of upper band and lower-band are $\pm 2$ and 0 for the dispersionless middle band at fixed $k_z$. In contrast, the triplon bands in region-I are topologically trivial. The region-I and region-II also appear in the reference Ref.\,\cite{triplon} without interlayer DMI $D_z=0$. In absence of interlayer interaction, the band topological transition in the low-lying excitation spectrum from region-I to region-II  occurs at a critical magnetic field $h_c=\frac{2\left|D_\perp^\prime\right|}{g_z}=1.316$T.

The nature of Weyl points in the remaining regions III, IV, V and VI depend on the sign of the DM-interaction $D'_\perp$. Here we describe the phase diagram for $D'_\perp<0$. The excitation spectrum for parameters in region-III is marked by two pairs of Weyl points at positions $(0,\pi,\pm k_{z_1})$ and $(\pi,0,\pm k_{z_1})$ as shown in Fig\,\ref{Weyl}(b), whereas region-IV features one-pair of Weyl points at $(0,0,\pm k_{z_2})$ (Fig.\,\ref{Weyl}(c)).
The number of Weyl points increases to three-pairs in region-V, located at momenta $(0,\pi,\pm k_{z_1})$, $(\pi,0,\pm k_{z_1})$ and $(0,0,\pm k_{z_2})$ (Fig.\,\ref{Weyl}(d)). Finally the triplon spectrum in the parameter regime region-VI features four-pairs of Weyl points at $(\pi,\pi,\pm k_{z_3})$, $(0,\pi,\pm k_{z_1})$, $(\pi,0,\pm k_{z_1})$ and $(0,0,\pm k_{z_2})$ as shown in Fig.\,\ref{Weyl}(e)-(f). The Weyl points at different sub-regions a, b, c, d in Fig.\,\ref{Phase}(a) are at the same position but, the charges of the Weyl points change. For the case $D'_\perp>0$, the Weyl nodes at $(0,0,\pm k_{z2})$ are substituted by the Weyl nodes at $(\pi,\pi,\pm k_{z3})$ and vice-versa. The results are summarized in Table\,\ref{Phase}(b).

\begin{widetext} 

\begin{figure}[htb]
\includegraphics[width=\textwidth]{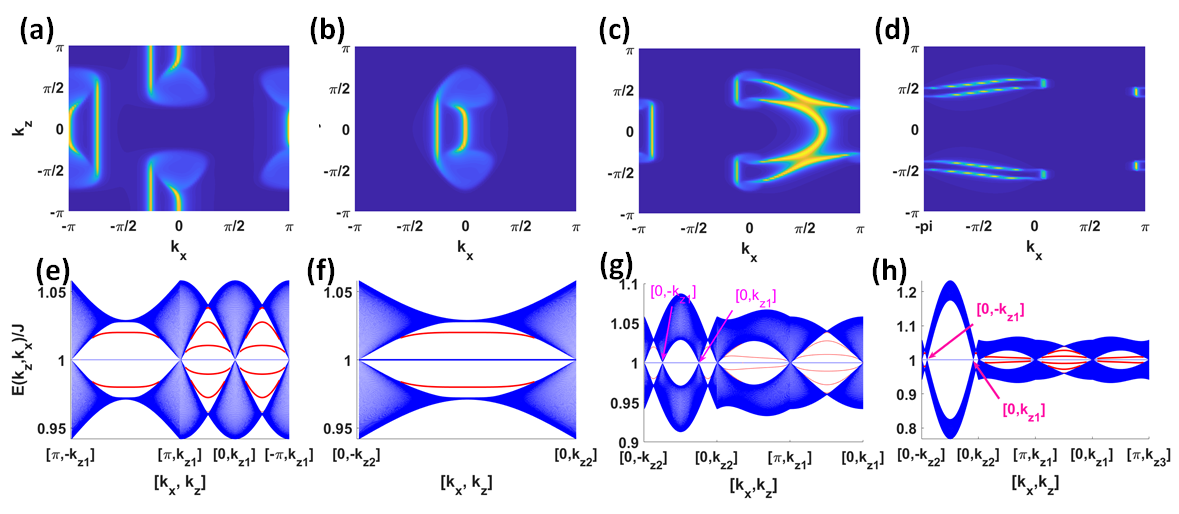} 
\caption{ Triplon-arcs on the $x$-$z$ surface in figures (a), (b), (c), (d)  for the parameters same as in (b), (c), (d), (e) in Fig.\,\ref{Weyl} respectively. The band structure of the system extended along $x$-$z$ direction in figures (e), (f), (g), (h) for the parameters same as in (a), (b), (c), (d) respectively. The surface states are shown in red-color. }
\label{Surface}
\end{figure}

\end{widetext}

\subsection{\label{sec3b} Surface arcs and surface states.}
The topological nature of the system drives the appearance of surface states in the material due to bulk-edge correspondence. In Fig.\,\ref{Surface}(a)-(d), we plot the surface spectral function of an infinite slab with periodic boundary condition along $x$-$z$ direction and open boundary condition along $y$-direction at the energy close to the energy of edge states as shown by the red lines in Fig.\,\ref{Surface}(e)-(h). 
The surface spectral function at energy $\omega$ and at $(k_x, k_z)$-point in the Brillouin zone is calculated as,
\begin{equation}
    \pazocal{A}^S(\omega,k_x, k_z)=-\frac{1}{\pi}\text{Imag}
    \left[\sum_n\frac{\pazocal{P}^S_n(k_x,k_z)}{\omega-E_n(k_x,k_z)+i\eta}\right],
\end{equation}
where $\pazocal{P}_n^S(k_x,k_y)$ and $E_n(k_x,k_y)$ are the probability at the surface and the energy respectively for the $n$-th eigenstate at $(k_x,k_z)$-point in Brillouin zone. 
 $\eta$ is a small positive number and we choose $\eta=10^{-3}$ for the numerical simulation.
Each of the projected bulk Weyl points on the surface emits two triplon-arcs, which indicate that the monopole charge of a Weyl point is $\pm 2$ satisfying the analysis about charges based on Chern number of bulk bands in previous subsection Sec.\,\ref{sec3a}. The surface triplon-arcs of the system have distinct characteristics in the different regions of parameter space, because of the different position and different numbers of Weyl points present in different sector in the parameter phase. The Fig.\,\ref{Surface}(a), (b), (c), (d) illustrates the surface triplon-arcs for the topological phase regime III, IV, V, VI respectively. For illustration, we describe the Fig.\,\ref{Surface}(a), which corresponds to the region-IIIa in phase diagram Fig.\,\ref{Phase}. There are two-pairs of Weyl triplon in this region, at positions $(0,\pi,\pm k_{z1})$ and $(\pi,0,\pm k_{z1})$. So the projected Weyl point on the $k_x$-$k_z$ surface exists at the positions $(\pi,\pm k_{z1})$ and $(0, \pm k_{z1})$. The pair of points along $k_z$ axis is connected by two surface triplon-arcs. The existence of surface triplon-arcs in the system can be detected using inelastic neutron scattering. The Fig.\,\ref{Surface}(e)-(h), describes that the surface states are chiral gapless state present within the bulk gap in the system.

\subsection{\label{sec3c} Thermal Hall effect for experimental detection.}  
Thermal Hall effect is the key experimental signature to detect topological excitations in a magnetic system.
In the past studies, the thermal Hall conductance was calculated for the topologically trivial and non-trivial gapped triplon bands for the two-dimensional counterpart of the model\,\cite{triplon,triplon2}.
The characteristic features of thermal Hall conductance of an Weyl triplon is different from the usual gapped topological triplon bands, making it an ideal probe to detect Weyl points. We calculate the thermal Hall effect in different regimes with  Weyl points (regimes III, IV, V or VI in Fig.\,\ref{Phase}(a)), gapped
topological triplons (regime II in Fig.\,\ref{Phase}(a)) and
gapped topologically trivial triplon excitations (regime I in Fig.\,\ref{Phase}(a)) to show that the 
thermal Hall conductivity exhibits distinct features identifying the 
different regimes. Since the Weyl points in this system always occur in 
pairs aligned along the $z$-direction, a transverse current cannot 
be created along the $z$-axis. Similarly, a temperature gradient along
this direction cannot produce a transverse current along any other direction\,\cite{WM1}. However, a transverse triplon current can be induced in $y$ (or $x$)-direction by applying a temperature gradient along the $x$ (or $y$)-direction. The thermal Hall conductance of the quasi-2D system with an applied field normal to the 2D planes is given by\,\cite{HallEffect1,HallEffect2,HallEffect3,HallEffect4},
\begin{align}
	 \kappa_{xy}&=\int^{\pi}_{-\pi} \frac{dk_z}{2\pi}\kappa_{xy}^{2D}(k_z),
\end{align}
where $\kappa_{xy}$ is the thermal Hall conductance and $\kappa_{xy}^{2D}(k_z)$ is the 2D-thermal Hall conductance contribution for the $k_x-k_y$-plane of fixed $k_z$-value in the Brillouin Zone, which is given by,
\begin{align}
    \kappa_{xy}^{2D}(k_z)=&\nonumber\\
    - T\int^\pi_{-\pi} &\int^\pi_{-\pi} \frac{dk_x dk_y}{(2\pi)^2} \sum_{n=1}^{N} \left[c_2(f^B[E_n(\bold{k})])-\frac{\pi^2}{3}\right] \Omega^z_n(\bold{k}),
\end{align}
where, $\bold{k}=(k_x, k_y, k_z)$ and $c_2(x)=(1+x)(\ln\frac{1+x}{x})^2-(\ln x)^2-2\text{Li}_2(-x)$, with $\text{Li}_2(x)$ as bilogarithmic function. 
The Berry curvature $\Omega^z_n(\bk)$ is defined in Eq.\,\ref{BerryCurvature}.
Furthermore $T$ denotes the temperature.
In general, the $\frac{\pi^2}{3}$-part of the expression does not contribute to the thermal Hall conductance, because the total Berry curvature summed over all bands and all $k$-points is zero.
For simplicity the physical quantities defined in the main text are dimensionless.
The experimentally measured physical quantities are connected to it's dimensionless counterpart by an multiplication factor as explained in the Appendix.\,\ref{appendixA}.


\begin{figure}[htb]
\includegraphics[width=0.5\textwidth]{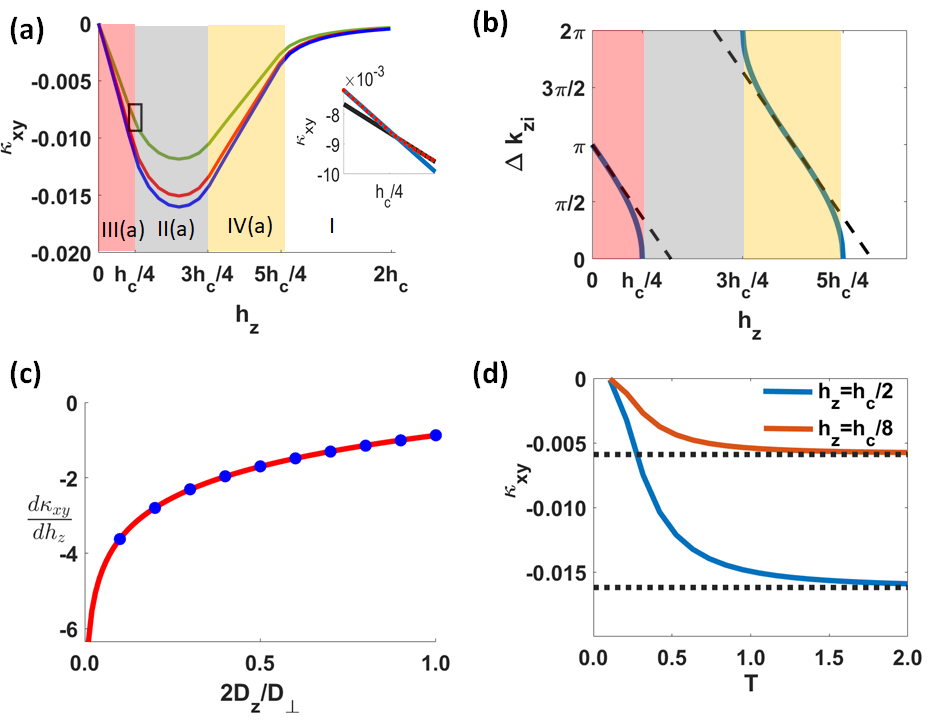} 
\caption{(a) The thermal Hall conductance as a function of the magnetic field is shown for $D_z=\frac{D'_\perp}{4}$ at different temperatures. The green, red and blue plots are for the temperatures $T$=0.5, 1.0 and 2.0 (\,$T^\prime$=17.33\,K, 34.66\,K, 69.32\,K\,) respectively. As magnetic field increases the system undergoes different phase regions as in Fig.\,\ref{Phase}(a). Inset of the figure shows the magnified region within the rectangular selection, which depicts that the tangent is undefined at the phase transition point. The red dots denote numerical data points. The blue and black lines denote the fitted data points using linear regression at the left and right side of phase transition point $h_z=\frac{h_c}{4}$ respectively. Moreover, the blue and black lines are further extended to the right and left side of point $h_z=\frac{h_c}{4}$ respectively to demonstrate inequality of the slopes. (b) The blue curves show the distance between Weyl points $\Delta k_{zi}$ vs the magnetic field $h_z$ for the Weyl triplon regions IIIa and IVa. The dotted black curve is the tangent of the curves at the point of inflection.
(c) The slope of plot in figure-(a) near $h_z\rightarrow 0$ region as a function of interlayer DMI $D_z$ at temperature $T=1$\,($T^\prime=10.4$K) is fitted with the fitting function $\frac{d\kappa_{xy}}{dh_z}=A\log(B|D_z|)+C$, where $A,\,B$ and $C$ are parameters.
 (d) The thermal Hall conductance as a function of temperature at different magnetic fields. The black dotted line denotes the maximum thermal Hall conductance achievable as the temperature is increased.}
\label{THE}
\end{figure}


While the nature and magnitude of interlayer DM-interaction ($D_z$) 
in \ce{SrCu2(BO3)2} has not been determined experimentally,
it is reasonable to expect finite $D_z$ as its presence is allowed by 
symmetry of the lattice. We assume a small, but finite, symmetry-allowed interlayer
DMI perpendicular to the layers. For a representative value of  $D_z=\frac{D'_\perp}{4}$, 
 the triplon bands lie in region IIIa of Fig.\,\ref{Phase}. The associated thermal Hall conductivity is plotted as a function of magnetic field in Fig.\,\ref{THE}(a)
for $0 \le h_z \le 2h_c)$. The  topology of the triplon band structure   undergoes several transitions in this range of applied field -- $\text{IIIa} \rightarrow \text{IIa} \rightarrow \text{IVa} \rightarrow \text{I}$ in Fig.\,\ref{Phase}. 
The excitation spectrum in region IIIa contains two pairs of Weyl points at $(0,\pi,\pm k_{z1})$ and $(\pi,0,\pm k_{z1})$, while there is one pair 
of Weyl points at $(0,0,\pm k_{z2})$ in the region IVa. The triplon
bands are fully gapped and topological in nature in region IIa,
whereas they are gapped and topologically trivial in region I. It is noted that, although the Berry curvature at a Weyl point is ill-defined, the thermal Hall conductance $\kappa_{xy}^{2D}(k_z)$ is a continuous function of $k_z$, because the right hand limit $k_z\rightarrow k_{zi}+0^+ $ and  the left hand limit as $k_z \rightarrow k_{zi}+0^- $ are equal, where $k_{zi}$ denotes the position of the Weyl point.
The thermal Hall conductivity depends on the Berry curvature distribution of the bands as well as the position of the Weyl points in Brillouin zone.

 In region-IIIa the thermal Hall conductance depends on the positions of the Weyl points.
At zero magnetic field the two pairs of Weyl points are located at $k_z=\pm \frac{\pi}{2}$, dividing the band structure into two different topological regions as in figure Fig.\,\ref{Weyl}(b).
The Berry curvature distribution of the bands in $k_x-k_y$ plane in between $k_z=\pm\pi$ and $k_z=\pm \frac{\pi}{2}$ is similar but opposite in sign compared to the bands in region between $k_z=-\frac{\pi}{2}$ and $k_z=+\frac{\pi}{2}$.
The contribution from the two regions cancel each other and the net thermal Hall conductivity is zero at zero magnetic field.
As the magnetic field increases the two pairs of Weyl points shifts towards the $k_z=0$ with constant $k_x-k_y$ coordinate and so the magnitude of thermal Hall conductivity increases due to inequality of the region with the opposite Berry curvature distribution.
Finally at magnetic field $h_z=\frac{h_c}{4}$ two pairs of Weyl points annihilate at $k_z=0$, creating fully gapped topological bands as in region IIa.
In the region IIa the Berry curvature distribution of all the bands in $k_x-k_y$ plane at different $k_z$ values are similar such that the Chern numbers of the lowest bands and upper-bands are $-2$ and $+2$ respectively.
The thermal Hall conductivity in this region increases (decreases) as a function of magnetic field before (after) $h_z=\frac{h_c}{2}$ because the magnitude of Berry curvature ($\Omega^z_n(\bk)$) of the lowest bands increases at lower (higher) energy values.
At magnetic field $h_z=\frac{3h_c}{4}$ the system enters region IVa where one pair of Weyl points appear at $k_z=\pm\pi$-plane.
The Weyl point in region-IVa divides the Brillouin zone into two parts along $k_z$-direction as in the figure Fig.\,\ref{Weyl}(c), one part of the Brillouin zone contains the bands in $k_x-k_y$ plane which are topologically trivial and another zone contains the bands in $k_x-k_y$ plane which are topologically non-trivial.
The Weyl points shifts towards $k_z=0$ plane with increase in magnetic field as a result the number of topologically non-trivial bands in plane $k_x-k_y$ plane decreases and so the thermal Hall conductivity decreases.
At magnetic field $h_z=\frac{5h_c}{4}$ the pair of Weyl points annihilate at the $k_z=0$-plane and the three bands become fully gapped (region-I).
In this region, the bands are topologically trivial with vanishing Chern numbers for individual bands in $k_x-k_y$ plane at a fixed $k_z$.
However the thermal Hall conductance in this region is still non-zero due to non-zero Berry curvature of the bands.
Finally at high enough magnetic field, thermal Hall conductance vanishes due to vanishing Berry curvature of the bands.
The slope of thermal conductivity in the Weyl triplon regime, as well as the second derivative of thermal Hall conductance near the phase transition, exhibit unique characteristics that are elaborately analyzed in this study.

 It is theoretically proposed that the presence of divergence in the derivative of thermal Hall conductance is a crucial signature of band-topological phase transition \,\cite{TopologicalMagnon2,trimerized,DerivativeThermalHall}.
To verify the presence of divergence in derivative of $\kappa_{xy}$ near phase transition point, we focus on the numerical data in the vicinity of the transition (marked by the rectangular box) in Fig.\,\ref{THE}(a). The data points on both sides of the transition exhibit a linear dependence on the applied field, with a discontinuous change in slope at the boundary, as shown in the inset of Fig.\,\ref{THE}(a).
Thus the tangent of the curve $\kappa_{xy}$ vs. $h_z$ is ill-defined at the phase transition point and as a result the double derivative of thermal Hall conductivity is divergent at the phase transition point.
Furthermore using simplistic model describing such band topological phase transition, we found that the double derivative $\frac{d^2\kappa_{xy}}{dh_z^2}\propto \log|h_z-h_p|$ logarithmic divergent in nature, where $h_p$ is the phase transition point (see Appendix.\,\ref{appendixB} and figure Fig.\,\ref{fig::Fitted}).

The thermal Hall conductance exhibits a unique linear 
dependence as a function of magnetic field for a region with Weyl points,
quite different from the phase region without Weyl point.
In reference Ref.\,\cite{HallConductance}, it is shown that the electronic Hall conductivity is proportional to the distance between Weyl points.
This feature is also observed in magnonic Weyl node systems\,\cite{WM1}.
In the regions IIIa and IVa the calculated thermal Hall conductivity (Fig.\,\ref{THE}(a)) is
in excellent agreement with a linear regression of the form 
$\kappa_{xy}=P\Delta k_{zi}+ Q$, where $P$, $Q$ are fitting parameters and $\Delta k_{zi}$ is the distance between the Weyl points.
The distance between the pair of Weyl nodes in the two regions are given by $\Delta k_{z1}(h_z)=2\cos^{-1}(-\frac{h_z g_z}{2D_z})$  and $\Delta k_{z2}(h_z)=2\cos^{-1}(-\frac{2D'_\perp+h_z g_z}{2D_z})$ respectively. The figure Fig.\,\ref{THE}(b) shows $\Delta k_{zi}$ as a function of magnetic field where it is clear that the curve is linear near the point of inflections $h_z=0$ and $h_z=h_c$. This yields the observed quasi-linear field dependence of the thermal Hall conductance in the Weyl triplon regions.
To summarize, the quasi-linear dependence thermal Hall conductance on applied field strength can serve as an experimental signature for the presence of the Weyl nodes.

The gradient of linear field dependence of $\kappa_{xy}$ in the Weyl triplon regions (region IIIa, IVa) depends strongly on the strength of inter-layer DMI. In figure Fig.\,\ref{THE}(c) we present the results for the calculated gradient
of $\kappa_{xy}$ as $h_z\rightarrow 0$. 
The magnitude of slope $\frac{d\kappa_{xy}}{dh_z}$ increases as the DMI $D_z$ decreases according to the figure Fig.\,\ref{THE}(c).
The slope is fitted as a function of $D_z$ using following fitting function $\frac{d\kappa_{xy}}{dh_z}=A\log(B|D_z|)+C$, where $A,\,B$ and $C$ are parameters. Thus in the absence of DMI $D_z$ the slope is infinite and so the plot $\kappa_{xy}$ against $h_z$ would cut the $h_z$-axis perpendicularly which is observed as in reference Ref.\,\cite{triplon}.
However in presence of interlayer DMI $D_z$ the plot of $\kappa_{xy}$ as a function of $h_z$ near $h_z\rightarrow 0$ region has a finite slope.
 The  infinite slope at $h_z\rightarrow 0$ for $D_z=0$ 
can be explained as follows. $h_z=0$ denotes a band topological transition point in absence of interlayer coupling~\,\cite{triplon} and gives rise to a divergence in the derivative of thermal Hall conductance\,\cite{DerivativeThermalHall,trimerized,TopologicalMagnon2}.
In presence of interlayer DMI $D_z$, $h_z=0$ is no longer a band topological transition point and as a result the slope become finite.
 Experimental measurement of the finite slope in the plot of thermal Hall conductance against magnetic field will reveal the magnitude of the interlayer DMI $D_z$.

 The temperature dependence of the thermal Hall conductivity is shown in the figure Fig.\,\ref{THE}(d). The magnitude of thermal Hall conductance increases with the temperature due to increase in thermally excited triplon density. The magnitude of variation in the value of $\kappa_{xy}$ with $T$ is greater in the Weyl triplon region IIIa (at $h_z=h_c/8$) compared to that in region IIa (at $h_z=h_c/2$) where the bands are fully  gapped. However the qualitative nature of the $\kappa_{xy}$ as a function temperature in two regions are similar.    
  At high temperature the thermal Hall conductivity is temperature-independent and attains it's maximum value\,\cite{ConstantTHE},
\begin{align}
    &\kappa_{xy}^{\text{max}} =\int_{-\pi}^{\pi} \frac{dk_z}{2\pi} \kappa_{xy}^{2D}(k_z) \nonumber\\
    &\text{where, }
    \kappa_{xy}^{2D\text{,max}}(k_z)=
    \int_{-\pi}^{\pi}\int_{-\pi}^{\pi}
    \frac{dk_x dk_y}{(2\pi)^2} \sum_{n=1}^{N}
    E_n(\bk) \Omega_n^z(\bk).
    \label{eq::HighTempTHE}
\end{align} 
The maximum achievable value of thermal Hall conductance is shown by the dotted line in Fig.\,\ref{THE}(d). 
Our treatment of the system is limited to quadratic triplon Hamiltonian in absence of interaction terms, but at high temperature due to higher triplon population the triplon-triplon interaction become significant.
So further study is required to investigate the temperature dependence of thermal Hall conductivity at high temperature taking into account the interaction terms.
 
\section{\label{sec4} Conclusion}
In conclusion we have demonstrated that \ce{SrCu2(BO3)2} is a possible host of Weyl triplons. Our study shows that interlayer perpendicular DMI (even if very small in magnitude) naturally give rise to the Weyl triplons. Furthermore the nature of triplon bands at low temperature  depends neither on the interlayer Heisenberg interation (because of orthogonal Dimer arrangement) nor on the interlayer in-plane DMIs, which makes the appearance of Weyl nodes robust against small deviations from the idealized model. Finally, We have shown that a finite slope of thermal Hall conductance as a function of magnetic field at $h_z \rightarrow 0$ region as well as the divergence in double derivative of thermal Hall conductivity near the phase transition from Weyl point region to topologically gapped band region, which are possible experimental signatures to detect the presence of interlayer DMI as well as Weyl nodes.

Inelastic neutron scattering provides an alternative way to probe Weyl nodes in triplon bands. Recent neutron scattering results for
\ce{SrCu2(BO3)2} show that exchange anisotropy and triplon bound 
states, neglected in this study, play 
an important role in determining the nature of the triplon bands\,\cite{triplon3}. 
Hybridization with the bound states lift the 3-fold degeneracy of the spin-1 Dirac point and consequently suppresses the appearance of a spin-1 Weyl point.
However, band-topological transitions still exist as a function of magnetic field and the band crossings are
expected to translate into a Weyl point in the 3D model in presence of interlayer DMI.
For the future, further theoretical and experimental study is required to investigate the presence and nature of Weyl triplons in real material.

\section*{ACKNOWLEDGEMENT}
Financial support from the Ministry of Education, Singapore, in the form
of grant MOE2018-T1-1-021 is gratefully acknowledged.

\pagebreak
\appendix
\begin{widetext}

\section{\label{appendixA} The connection between dimensionless and real physical quantities}
In the main text we have calculated the dimensionless Physical quantities and in this appendix we show the dimensionless physical quantities are connected to the experimentally measured physical quantities through a multiplication factor.
The notations used for the dimensionless physical quantities are unprimed, whereas the notations used for the experimentally measured physical quantities are primed.

The expression of thermal Hall conductivity for a two-dimensional material is given by\,\cite{HallEffect4},
\begin{equation}
\kappa_{xy}^{\prime 2D}(k_z^\prime)=\frac{2k_B^2 T'}{\hbar A} \sum_{\bk}\sum_{n=1}^N \left\lbrace c_2[\rho_n]-\frac{\pi^2}{3}\right\rbrace \text{Im}\braket{\frac{\partial u_n(\bk)}{\partial k_x^\prime}}{\frac{\partial u_n(\bk)}{\partial k_y^\prime}},\label{Aeq1}
\end{equation}
 $(k_x^\prime,k_y,^\prime,k_z^\prime)$ are the components of crystal-momentum in the reciprocal-space and the first Brillouin zone is defined by $-\frac{\pi}{a_i}\leq k_i^\prime < \frac{\pi}{a_i}$ ($i\in x,y,z$ and $a_i$ is the lattice constant in $i$-th direction and $a_x=a,\,a_y=b,\,a_z=c$). 
In integral-form the equation Eq.\,\ref{Aeq1} is transformed as,
\begin{align}
\kappa_{xy}^{\prime 2D}(k_z^\prime)=
\frac{2k_B^2 T'}{\hbar } \frac{1}{(2\pi)^2}\sum_{n=1}^N \int^{\frac{\pi}{a}}_{-\frac{\pi}{a}}\int^{\frac{\pi}{b}}_{-\frac{\pi}{b}} dk_x^\prime dk_y^\prime 
\left\lbrace c_2[\rho_n]-\frac{\pi^2}{3}\right\rbrace \text{Im}\braket{\frac{\partial u_n(\bk)}{\partial k_x^\prime}}{\frac{\partial u_n(\bk)}{\partial k_y^\prime}}.
\label{Aeq2}
\end{align}
The dimensionless crystal-momentum is defined as $k_i=a_i k_i^\prime$. 

Using the transformation relations the equation eq.\,\ref{Aeq2} becomes,
\begin{align}
\kappa_{xy}^{\prime 2D}(k_z^\prime)
&=\frac{2k_B^2 T'}{\hbar } \frac{1}{(2\pi)^2}\sum_{n=1}^N \int^{\pi}_{-\pi}\int^{\pi}_{-\pi} dk_x dk_y 
\left\lbrace c_2[\rho_n]-\frac{\pi^2}{3}\right\rbrace \text{Im}\braket{\frac{\partial u_n(\bk)}{\partial k_x}}{\frac{\partial u_n(\bk)}{\partial k_y}}\nonumber\\
&=-\frac{k_B^2 T'}{\hbar } \frac{1}{(2\pi)^2}\sum_{n=1}^N \int^{\pi}_{-\pi}\int^{\pi}_{-\pi} dk_x dk_y 
\left\lbrace c_2[\rho_n]-\frac{\pi^2}{3}\right\rbrace \Omega_n^z(\bk).
\label{Aeq4}
\end{align}
 Considering $J=722GHz$ as energy unit the dimensionless temperature is defined as $T=\frac{k_B T^\prime}{J}$, where $T^\prime$ is the temperature in Kelvin. Thus the measured 2D conductivity in terms of it's dimensionless counterpart is given by,
\begin{align}
\kappa_{xy}^{\prime 2D}(k_z^\prime) &=\kappa_{xy}^{2D}(k_z^\prime) \frac{k_B J}{\hbar}
\label{eqeq}\\
\text{where,}\, \kappa_{xy}^{2D}(k_z^\prime) &=- \frac{T}{(2\pi)^2}\sum_{n=1}^N \int^{\pi}_{-\pi}\int^{\pi}_{-\pi} dk_x dk_y 
\left\lbrace c_2[\rho_n]-\frac{\pi^2}{3}\right\rbrace \Omega_n^z(\bk)
\end{align}
The dimensionless thermal Hall conductivity is given by,
\begin{align}
\kappa_{xy}&= \frac{1}{N_z}\sum_{k_z^\prime} \kappa_{xy}^{2D}(k_z^\prime)\nonumber\\
&=\frac{c}{2\pi} \int^{\pi/c}_{-\pi/c} dk_z^\prime \kappa_{xy}^{2D}(k_z^\prime)\nonumber\\
&=\frac{1}{2\pi} \int^\pi_{-\pi} dk_z \kappa_{xy}^{2D}(k_z),
\label{eqeq2}
\end{align}
where $N_z$ is the number of unit-cell along the $z$-direction. Using equation Eq.\,\ref{eqeq} and Eq.\,\ref{eqeq2}, the experimentally measured thermal Hall conductivity in terms of it's dimensionless counterpart is given by,
\begin{equation}
\kappa_{xy}^\prime=\kappa_{xy}\frac{N_zk_BJ}{\hbar}
\end{equation}

\section{\label{appendixB} Divergence of double derivative of thermal Hall conductivity at phase transition point}

\begin{figure}[htb]
\includegraphics[width=0.5\textwidth]{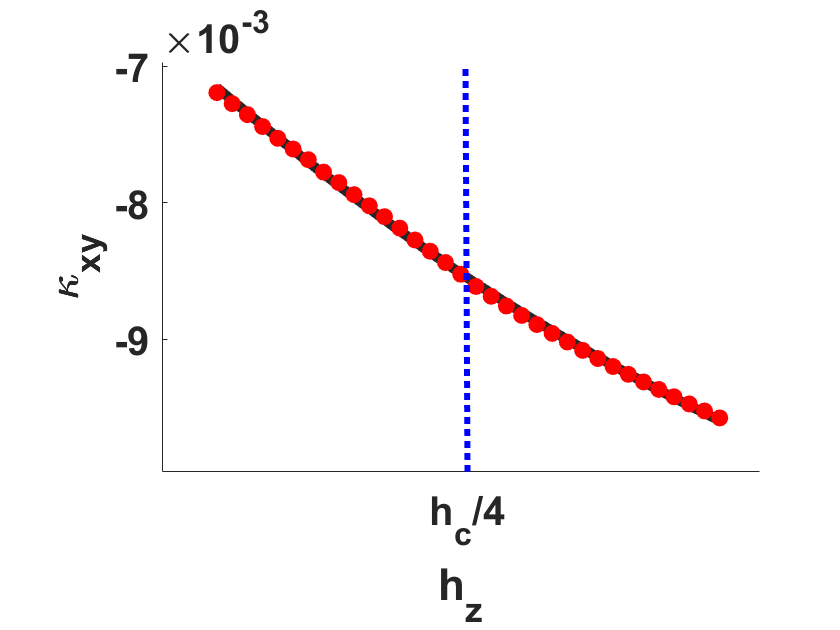} 
\caption{The plot of thermal Hall conductance as function of magnetic field within the rectangular box in figure Fig.\,\ref{THE}(a). The red dots denote the numerical data points and the black line represents fitting of that data using the expression in equation Eq.\,\ref{eq::B6}. The fitting parameters are $A=-25.78,\,B=-1.2181,\,C=-0.0086$.}
\label{fig::Fitted}
\end{figure}

We consider the following simplistic two band model which describes the band-topological phase transition from the region of Weyl points to a region of gapped topological bands,
\begin{align}
    \pazocal{H}&=k_x\sigma_x+k_y\sigma_y+\left|k_z-c\right|\sigma_z,\,\,\,k_z\geq 0
    \nonumber\\
    &=k_x\sigma_x+k_y\sigma_y+\left|k_z+c\right|\sigma_z,\,\,\,k_z<0.
\end{align}
If $c>0$, a pair of Weyl points exist at $k_z=\pm c$ and if $c<0$, the system consists of topologically gapped bands.
Thus $c=0$ is the band-topological phase transition point.
For simplicity, we use with the formula for thermal Hall conductance at high temperature (see Eq.\,\ref{eq::HighTempTHE}),
\begin{align}
    \kappa_{xy}&\propto\int\int d^2k dk_z \sum_n \Omega_n^z(\bk) E_n(\bk)
    \nonumber\\
    &\propto
    \int \int d^2k dk_z \Omega^z_1(\bk) E_1(\bk),\quad\left[\because \Omega_1(\bk)=-\Omega_2(\bk) \text{ and }E_1(\bk)=-E_2(\bk)\right]
\end{align}
where $k=\sqrt{k_x^2+k_y^2}$. The energy and the Berry curvature expressions for the lower band are given by,
\begin{equation}
    E_1(\bk)=-\sqrt{k^2+(k_z\mp c)^2}\,\,,\quad
    \Omega_1^z(\bk)=-\frac{2\left|k_z\mp c\right|}{\left((k_z\mp c)^2+k^2\right)^{3/2}},
\end{equation}
where the negative and positive sign in front of $c$ is for $k_z\geq 0$ and $k_z<0$ region respectively.
The thermal Hall conductivity is given by,
\begin{align}
    \kappa_{xy}&\propto\int_{k=0}^{k=k_c}
    \int_{k_z=-\pi}^{k_z=+\pi}
    d^2k dk_z
    \frac{\left|k_z\mp c\right|}{(k_z\mp c)^2+k^2}
    \nonumber\\
    &\propto
    \int_{k=0}^{k=k_c}
    \int_{k_z=0}^{k_z=+\pi}
    d^2k dk_z
    \frac{\left|k_z- c\right|}{(k_z- c)^2+k^2}
    \quad
    \left[\because \text{The integrand is an even function of }k_z\right]
    \nonumber\\
    &\propto
    \left[
    \left(k_c^2+(\pi-c)^2 \right)\log(k_c^2+(\pi-c)^2)-
    \left(k_c^2+(\pi-c)^2\right)
    \right]
    -\left[
    (\pi-c)^2\log(\pi-c)^2-(\pi-c)^2
    \right]
    \nonumber\\
    &\qquad\qquad\qquad\qquad\qquad\qquad\qquad
    -\left[
    \left(k_c^2+c^2\right)\log\left(k_c^2+c^2\right)
    -\left(k_c^2+c^2\right)
    \right]
    +\left[
    c^2\log c^2-c^2
    \right],
\end{align}
where we assume the Berry curvature is only important in the region $k\leq k_c$.
Near phase transition point $c\rightarrow 0$ we have $k_c \gg c$ and $k_c\ll \pi-c$,
\begin{align}
\kappa_{xy}
&\propto
c^2\log\left|c\right|
\nonumber\\
\therefore\frac{d^2\kappa_{xy}}{dc^2}
&\propto
\log\left|c\right|
\end{align}
Thus double derivative of thermal Hall conductivity with respect to $c$ is logarithmically divergent near phase transition.
For further validation, the plot within the rectangular-box in figure Fig.\,\ref{THE}(a) is fitted using the following expression,
\begin{equation}
    \kappa_{xy}=A\left(h_z-\frac{h_c}{4}\right)^2\log\left|h_z-\frac{h_c}{4}\right|+B\left(h_z-\frac{h_c}{4}\right)+C,
    \label{eq::B6}
\end{equation}
and shown in figure Fig.\,\ref{fig::Fitted}.
\end{widetext}

\bibliographystyle{apsrev4-1}

%

\end{document}